\documentclass[
    ,final            % use final for the camera ready runs
    ,mathptm                 % add further options here if necessary
  ]
  {aipproc}

\layoutstyle{6x9}

\begin{document}

\title{Review of Two-Photon Interactions}

\author{David Urner}{
  address={Cornell University}
}

\begin{abstract}
Presented are recent results of two-photon interactions. Topics inlcude photon structure functions, 
inclusive hadron production, differential cross sections derived from 
tagged $\gamma\gamma$ fusion events and results in exclusive hadron production, particularly the observations of
the $\eta_c^\prime$.
\end{abstract}

\maketitle

%%%%%%%%%%%%%%%%%%%%%%%%%%%%%%%%%%%%%%%%%%%%
%% MAINMATTER
%%%%%%%%%%%%%%%%%%%%%%%%%%%%%%%%%%%%%%%%%%%%
Two-photon interactions provide a unique opportunity to study a large variety 
of physics topics. In electron positron machines two photons are emitted producing 
the photon-photon interaction. One can use this process to examine the structure 
and interaction of the photon. Two-photon data allow for a large number of tests of 
perturbative QCD and the exploration of non-perturbative phenomena in 
the light meson sector. 

\section{Photon Structure Function}
Two photon interactions are instrumental to extract the hadronic structure function 
of the photon, which is related to the quark densities in the photon. It is measured 
in single tagged events, which means that one of the photons is real, and the other 
has a large momentum transfer much larger than $\Lambda_{QCD}$. This process can be viewed 
as deep inelastic electron e-$\gamma$ scattering, where a quasi real photon is probed 
by a virtual photon with high Q$^2$. The differential cross section $F_2^\gamma(x,Q^2)$ is related to the 
hadronic structure function: $\frac{d^2\sigma_{e\gamma\rightarrow eX}}{dxdQ^2} = [(1+1-y)^2)F_2^\gamma(x,Q^2)-y^2F_L^\gamma(x,Q^2)]\frac{2\pi\alpha^2}{xQ^2}$ with $x\sim\frac{Q^2}{Q^2+W^2},$ $y=\frac{E_{tag}}{E_{beam}} cos^2\theta$. The longitudinal term can usually be neglected since 
y$^2$ is small.
The LEP2 data has considerably extended the reach towards small x.
The present level of precision starts to challenge current structure function 
parameterizations. The newest result comes from ALEPH in the Q$^2$ range of 17.3 
and 67.2 GeV$^2$~\cite{aleph_structure}.

\section{Heavy Quark Production}
Heavy quark production, provides important tests of perturbative QCD as well as probing 
the partonic densities in the photon. Recent results include a D* measurement of ALEPH ~\cite{aleph_heavy_quark}
and a muon semileptonic  measurement by DELPHI~\cite{delphi_heavy_quark}. ALEPH shows 
that besides W, also the transverse momentum and the differential cross section in 
p$_T$ and $\eta$ are in good agreement with NLO QCD. 
DELPHI added a preliminary third measured point of the total bottom cross section, which agrees well with 
results from OPAL and L3. The large quark mass should allow good 
accuracy in perturbative calculations. However the current status of NLO QCD underestimates 
the cross sections by about a factor of 3. This discrepancy cannot be tuned away by 
changing the b-quark mass.

\section{Inclusive Meson Production}
L3 has complemented their earlier result in inclusive $\pi^0$ production with an inclusive 
charged pion and kaon production measurement~\cite{l3_inclusive_pion}, using the LEP2 data in the region of 
$W_{\gamma\gamma}<$5 GeV and Q$^2<$8GeV$^2$, see Figure~\ref{fig:ppbar}a). 
There is a good agreement between data and NLO QCD below a p$_T$ of 3 GeV, but above about 5 GeV 
the data clearly surpass the expectation similar to the excess previously observed in the $\pi^0$ and $K_s$ spectra. 
In the inclusive single jet production the data is compared to a NLO QCD calculation~\cite{citbertora}, 
which agrees well with many inclusive observables for the older OPAL data. 
However for the L3 data, see figure~\ref{fig:ppbar}b), which go to a higher $p_T$, 
there is again a clear deviation observed~\cite{l3_jet}.

\section{Exclusive hadron pair production}
For small photon virtualities, large s and large momentum transfer from 
the photons to the hadrons, treating the $\gamma\gamma\rightarrow hh$ 
system in leading twist perturbation theory, the
transition amplitude factorizes into a hard scattering amplitude 
$\gamma\gamma\rightarrow q \overline{q} q \overline{q}$ + a single hadron 
distribution for each hadron~\cite{citbrodskylepage}. The hand bag model
adds a soft 2-hadron distribution amplitude in what is basically a power 
correction.

\subsection{Untagged $\gamma\gamma\rightarrow\pi\pi,KK$, Baryon-Antibaryon}
The handbag model predicts the ratio 
$\frac{\gamma\gamma\rightarrow\pi\pi}{\gamma\gamma\rightarrow KK}$=1~\cite{citdiehl_pipi}, while
the leading twist calculation would predict a ratio of 2. Aleph, Delphi~\cite{cit2pi_aleph_delphi},
and with higher statistics, but still preliminary, BELLE~\cite{cit2pi_belle} have measured 
these decays and find indeed that the ratio is around 1 over the 
full W range. The handbag model also predicts a 1/sin$^4$($\theta$)
behavior for the differential cross section, which agrees well with
the data.

\begin{figure}[!b]
 \resizebox{0.98\columnwidth}{!}
  {
    \resizebox{0.3\columnwidth}{!}
      {
        \includegraphics{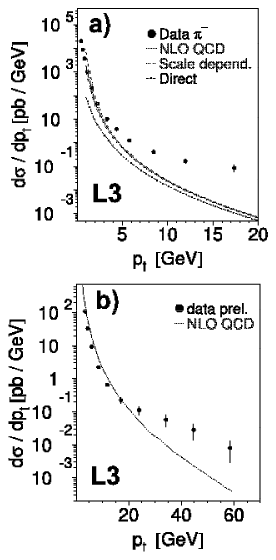}	
      }
    \resizebox{.6\columnwidth}{!}
      {
	\includegraphics{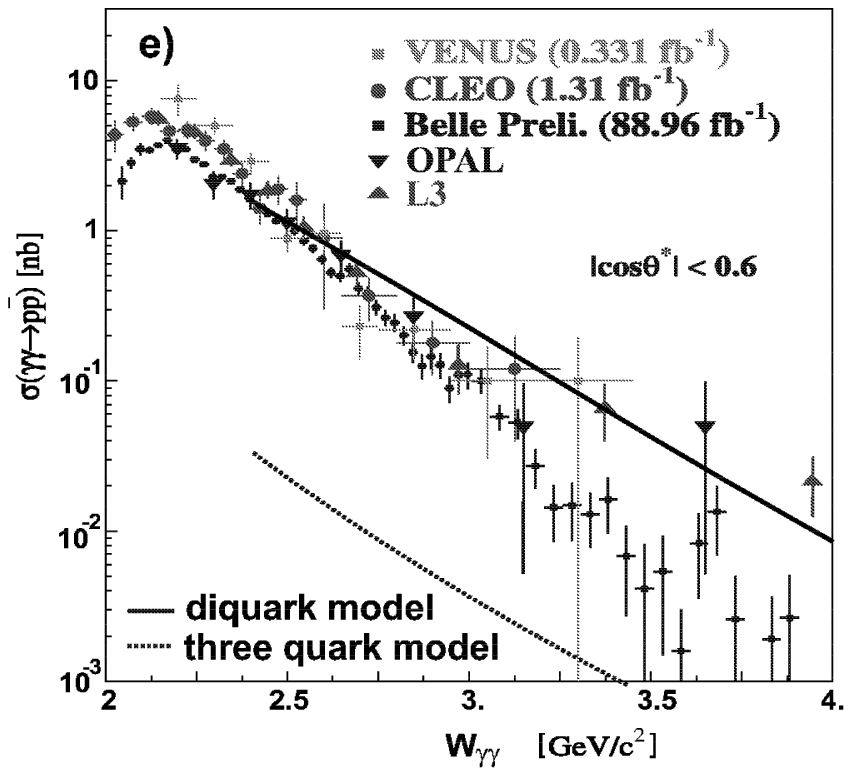}
      }
    \resizebox{0.29\columnwidth}{!}
      {
	\includegraphics{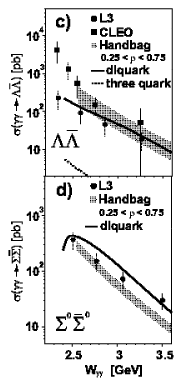}
      }
  }
 \caption{L3 experiment: Inclusive $\pi^-$ (a) and single jet production (b) cross section. 
                         Exclusive $\gamma\gamma\rightarrow \Lambda\Lambda$ (c) and $\gamma\gamma\rightarrow \Sigma\Sigma$ (d) 
			 cross section and comparison with handbag and diquark models.\newline
			 BELLE experiment: Exclusive $\gamma\gamma \rightarrow p \overline{p}$ cross section (e) 
 			 and comparison with diquark model.}
 \label{fig:ppbar}
\end{figure}

Comparing predictions made in the framework of the hard scattering approach~\cite{citbrodskylepage} with
the $\gamma\gamma\rightarrow$ Baryon-Antibaryon cross section one is sensitive to the quark structure of
the baryon. Reasonable agreement with the quark-diquark model~\cite{diquark} is found by the L3, OPAL, CLEO and 
BELLE~\cite{citppbar_exp}
experiments in the $p\overline{p}$, $\Lambda\overline{\Lambda}$ and $\Sigma\overline{\Sigma}$ final states,
see figures~\ref{fig:ppbar}c)-\ref{fig:ppbar}e), although the BELLE data starts to challenge the diquark model,
while three-quark model~\cite{threequark} predictions are too low.
The $\gamma\gamma\rightarrow p\overline{p}$ spectra can be used to fit the parameters of the handbag model, and hence
make predictions for the cross sections of all other baryon octet members~\cite{diehl_baryon}, using one additional 
parameter $\rho$ - a ratio of form factors of the proton. They agree with the
$\Lambda\overline{\Lambda}$ and $\Sigma\overline{\Sigma}$ measurement by CLEO and L3~\cite{citlambda}, 
see figure~\ref{fig:ppbar}c),~\ref{fig:ppbar}d).
One notes that for $W_{\gamma\gamma}$ below 2.6 GeV there are experimental discrepancies between the 
BELLE and CLEO experiments for the $p\overline{p}$~\ref{fig:ppbar}e) and between the L3 and CLEO experiments for the 
$\Lambda\overline{\Lambda}$~\ref{fig:ppbar}c) final states, which should be resolved.

\subsection{$\gamma\gamma\rightarrow\rho\rho$}

The L3 experiment measured $\gamma\gamma\rightarrow\rho^0\rho^0$ and $\gamma\gamma\rightarrow\rho^+\rho^-$ 
processes~\cite{citrho}. A simple partial wave analysis was performed on the 4-pion final states, including only $\rho\rho$ 
partial waves plus a 4 $\pi$ isotropic background, fitting the data separately for each W bin. 
The only contributing waves have a $J^P,J_Z$ of $0^+$ and $2^+,2$ as shown in figure~\ref{fig:rhorho}. 
From this result one would conclude, that the isospin ratio of the $\rho\rho$ cross sections is incompatible 
with either I=0 or I=1.

The L3 experiment also analyzed the two-photon production of $\rho^0\rho^0$ in the single tagged mode at $\sqrt{s}$
= 89-209 GeV~\cite{citrho}, which allow a test of the qq, gg $\rightarrow$ meson-pair mechanism. The $\rho^0\rho^0$ 
signal was separated from the $\rho^0\pi^+\pi^-$ and 4$\pi$ backgrounds with the box method. 
Figure~\ref{fig:rhorho}c) shows the $\gamma^*\gamma$ differential cross section as a function of $Q^2$, which agrees well with
the generalized vector dominance model.

\subsection{Single Tagged $\gamma^*\gamma\rightarrow\pi^0\pi^0$}
\begin{figure}[t]
 \resizebox{1.0\columnwidth}{!}
  {
    \resizebox{.65\columnwidth}{!}
      { 
        \includegraphics{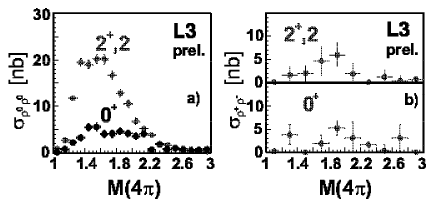}
      }
    \resizebox{.33\columnwidth}{!}
      { 
	\raisebox{0.4cm}
	  {
            \includegraphics{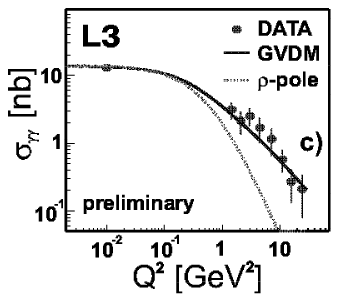}
          }
      }
    \resizebox{.28\columnwidth}{!}
      {
        \includegraphics{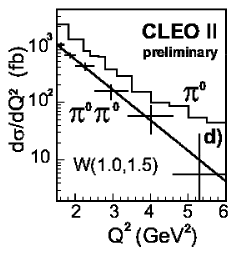}
      }
  }
 \caption{L3 experiment: cross section of contributing partial waves to untagged $\gamma\gamma\rightarrow \rho^0\rho^0$ (a) 
                         and $\gamma\gamma\rightarrow \rho^+\rho^-$ (b) decays. Single tagged $\gamma^*\gamma\rightarrow\rho^0\rho^0$
			 cross section (c). \newline
	  CLEO experiment: Single tagged $\gamma^*\gamma\rightarrow\pi^0\pi^0$ cross section 
			   compared to $\gamma^*\gamma\rightarrow\pi^0$ (d).}
 \label{fig:rhorho}
\end{figure}

In tagged two-photon decays the process $\gamma^*\gamma\rightarrow\pi\pi$ in the region of large $Q^2$ but small 
$W$ factorizes into a perturbatively calculable part dominated by short distance scattering~\cite{citbrodskylepage}: 
$\gamma^*\gamma\rightarrow q \overline{q}$ or, 
$\gamma^*\gamma\rightarrow g g$ and non perturbative matrix elements measuring the transitions 
$q \overline{q}\rightarrow\pi\pi$ and $q \overline{q}\rightarrow g g$, called generalized distribution 
amplitudes~\cite{diehl_tagged}. CLEO has a preliminary measurement of the $\gamma^*\gamma\rightarrow\pi^0\pi^0$ 
cross sections for different $Q^2$ and W bins shown in figure~\ref{fig:rhorho}d). 

\subsection{$\gamma\gamma\rightarrow\eta(1440)$}
	Until recently $\eta(1440)$ has only been seen in gluon rich environments such as $\overline{p}p$ 
annihilation or J/$\psi$ decay. Although quenched lattice calculations indicate a mass of the $0^{-+}$ glueball around
2GeV, some glueball content of the $\eta(1440)$ can presently not be excluded. The L3 experiment reported a 
first observation of the $\eta(1440)$ in two-photon collisions~\cite{citl3_eta1440} with a 
$\Gamma_{\gamma\gamma}(\eta_{1440})\cdot BR(\eta_{1440}\rightarrow K_s K^+ \pi^-)$ = 49 $\pm$ 12 eV.
Since two-photon partial width of glueballs should be very small, this would indicate that the $\eta(1440)$
is mostly not a glueball. 

\begin{figure}[!b]
 \resizebox{.7\columnwidth}{!}
  {
    \resizebox{.43\columnwidth}{!}
      { 
        \includegraphics{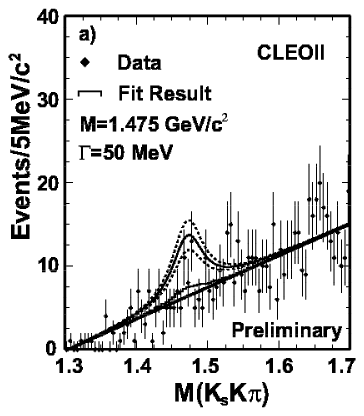}
      }
    \resizebox{.46\columnwidth}{!}
      { 
        \includegraphics{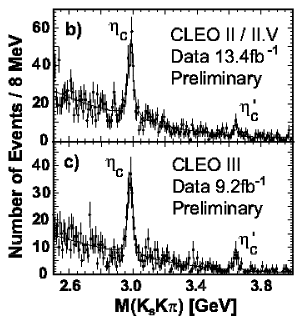}
      }
  }
 \caption{CLEO experiment: Search for $\gamma\gamma\rightarrow \eta(1440) (\eta(1440)\rightarrow K_s K \pi$ (a), 
			   the expected L3 signal (line) with errors (dashed) is superimposed. 
	                   Observation of $\eta_c^\prime$ decaying to $K_s K \pi$ final state for CLEO II (b) and 
		           CLEO III (c) data sets.}
 \label{fig:etacprime}
\end{figure}

CLEO has analyzed 13.8fb$^{-1}$ of data collected around the $\Upsilon (4S)$ energies and searched for 
$\gamma\gamma\rightarrow\eta(1440)\rightarrow\ K_s K^{\pm}\pi^{\mp}$ decays~\cite{citcleo_eta1440}. There is no $\eta(1440)$ resonance
observed and an upper limit for the two-photon partial width of 14.4 eV is obtained. This result, which
includes all systematic errors, is 2.9 $\sigma$ below the L3 result. Figure~\ref{fig:etacprime}a) shows the CLEO data with the 
fit result and overlayed the signal (line) with errors (dashed), as expected from the L3 result.

\section{$\eta_c^\prime$ Observation}
The total existing experimental knowledge of the hyperfine splitting in any meson system is
the $\Delta M$ = M(J/$\psi$)-M($\eta_c$) = 117 $\pm$ 2 MeV. The measurement of 
$\Delta M$ = M($\psi^\prime$)-M($\eta_c^\prime$) is important for the understanding of 
the spin-spin interaction in the confinement region. Theoretical predictions based on 
potential model calculations are M($\eta_c^\prime$) $\sim$ 3594-3629 MeV, with a two-photon 
partial width ratio of $\Gamma_{\gamma\gamma}(\eta_c^\prime)/\Gamma_{\gamma\gamma}(\eta_c)\sim$ 0.7.
A Crystal Ball measurement~\cite{citCrystalBall} at 3594 $\pm$ 5 MeV could not be confirmed by other searches~\cite{citNegative}. 
Recently the BELLE experiment published an $\eta_c^\prime$ observation in 
$B\rightarrow K(K_s K^{\mp}\pi^{\pm})$~\cite{citbelle_b_etacprime} with a mass of 3654$\pm$6$\pm$8 MeV and a significance 
of more than 6$\sigma$. The BELLE experiment also observed a $\eta_c^\prime$ signal in the mode 
$e^+e^-\rightarrow J/\psi(X)$~\cite{citbelle_jpsix} with a mass of 3622$\pm$12 MeV and a significance of 3.4$\sigma$.
The BABAR experiment has presented a preliminary $\eta_c^\prime$ signal~\cite{citbabar_etacprime} in two-photon decays with a mass of 
3633$\pm$5$\pm$1.8 MeV.

The CLEO experiment has analyzed 13.9fb$^{-1}$ taken with the CLEO II detector~\cite{citcleo_etacprime}, see figure~\ref{fig:etacprime}b) 
and finds a signal at a mass of 3642.7$\pm$4.0 MeV with a significance of 4$\sigma$, a significance which 
assumes the mass or width of the resonance not to be known. CLEO confirmed this observation using 9.2fb$^{-1}$
of data from the CLEOIII detector~\cite{citcleo_etacprime}, see figure~\ref{fig:etacprime}c), 
finding a mass of 3642.5$\pm$3.4 MeV 
(no systematic error included) and a significance of 5.7$\sigma$. The ratio 
$\frac{\Gamma_{\gamma\gamma}(\eta_c^\prime)xB(\eta_c^\prime\rightarrow 
K_sK\pi)}{\Gamma_{\gamma\gamma}(\eta_c)xB(\eta_c\rightarrow K_sK\pi)}=0.17\pm0.06$ (CLEO II) $0.29\pm0.09$ 
(CLEO III), with statistical errors only. All results are preliminary.

%%%%%%%%%%%%%%%%%%%%%%%%%%%%%%%%%%%%%%%%%%%%%%%%
%% BACKMATTER
%%%%%%%%%%%%%%%%%%%%%%%%%%%%%%%%%%%%%%%%%%%%%%%%

\end{document}